\documentclass{article}

\usepackage{arxiv}

\usepackage{mathtools}  
\mathtoolsset{showonlyrefs}  
\usepackage{xfrac}
\usepackage{amsthm,amssymb}
\numberwithin{equation}{section}
\newtheorem{example}{Example}
\usepackage{xcolor}
\usepackage{bm}
\usepackage{natbib}

\DeclareMathOperator{\sech}{sech}
\newcommand{\R}{\mathbb{R}}
\newcommand{\bfx}{\mathbf{x}}
\newcommand{\bfy}{\mathbf{y}}
\newcommand{\bfX}{\mathbf{X}}
\newcommand{\bfY}{\mathbf{Y}}
\newcommand{\bff}{\mathbf{f}}

\newtheorem{remark}{Remark}
\usepackage[linesnumbered,ruled]{algorithm2e}
\SetKwInput{KwInput}{Required}                
\SetKwInput{KwOutput}{Output}              
\RestyleAlgo{ruled}

\usepackage{url,hyperref,lineno,microtype,subcaption}
\usepackage[onehalfspacing]{setspace}

\title{Identification of Moment Equations via Data-Driven Approaches in Nonlinear Schr{\"o}dinger Models} 

\author{  Su Yang\\
    Department of Mathematics and Statistics\\
  University of Massachusetts Amherst\\
  Amherst, MA 01003,  USA \\
  \texttt{suyang@umass.edu} \\
   \And
    Shaoxuan Chen\\
    Department of Mathematics and Statistics\\
  University of Massachusetts Amherst\\
  Amherst, MA 01003,  USA \\
  \texttt{shaoxuanchen@umass.edu} \\
\And
    Wei Zhu\\
    School of Mathematics\\
    Georgia Institute of Technology\\
    Atlanta, GA 30332, USA\\
  \texttt{weizhu@gatech.edu} 
  \And
    P.G. Kevrekidis\\
    Department of Mathematics and Statistics\\
  University of Massachusetts Amherst\\
  Amherst, MA 01003,  USA \\
  \texttt{kevrekid@umass.edu} 
}

\begin{document}
\maketitle

\begin{abstract}

The moment quantities associated with the nonlinear
   Schr\"{o}dinger equation offer important insights
   towards the evolution dynamics of 
   such dispersive wave partial differential equation (PDE) models. The effective dynamics of the moment quantities is amenable to both analytical and numerical treatments. In this paper we present a data-driven approach associated with the ``Sparse Identification of Nonlinear Dynamics'' (SINDy) to numerically capture the evolution behaviors of such moment quantities. Our method is applied first to some well-known closed systems of ordinary differential equations (ODEs) which describe the evolution dynamics of relevant moment quantities. Our examples are, progressively, of
   increasing complexity and our findings 
   explore  different choices within the SINDy library.
   We also consider the potential discovery of 
   coordinate transformations that lead to 
   moment system closure.
   Finally, we extend considerations
   to settings where a closed analytical form
   of the moment dynamics is not available. 
  
\end{abstract}


\section{Introduction and Motivation}

The study of Nonlinear Schr{\"o}dinger type 
models~\citep{Sulem,AblowitzPrinariTrubatch}
is of wide interest and significance in a diverse
array of physical modeling settings~\citep{ablowitz:sne91,ablowitz2}.
Relevant areas of application extend
from atomic physics~\citep{becbook2,pethick,siambook}
to fluid mechanical problems~\citep{ablowitz2,infeld}
and from plasma physics~\citep{kono,infeld} to
nonlinear optics~\citep{Kivshar2003,hasegawa:sio95}.
Indeed, the relevant model is a prototypical
envelope wave equation that describes the
dynamics of dispersive waves. Specifically,
in the context of nonlinear optics, it
describes the envelope of the electric field
of light in optical fibers (as well as waveguides), 
with the relevant measurable quantity being
the light intensity $I$ proportional to the
square modulus of the complex field $u(x,t)$.
Generalizations of relevant optical applications,
involving multiple polarizations or frequencies
of light have also been widely considered, in both
spatially homogeneous and spatially heterogeneous 
media~\citep{Kivshar2003,hasegawa:sio95}.

At the same time, over the past few years, there
has been an explosion of interest in data-driven
methods, whereby machine-learning techniques
are brought to bear towards understanding,
codifying and deducing the fundamental
quantities of physical systems. 
Arguably, a turning point in this effort was
the development of the so-called physics-informed
neural networks (PINNs) by \citep{raissi_physics-informed_2019}
and of similar methodologies such as the extension
of PINNs via the so-called DeepXDE~\citep{lu2021deepxde},
as well as the parallel track of sparse identification
of nonlinear systems, so-called SINDy by \citep{Kutz_SINDy}, which is
central to the considerations herein. 
Additional methods include, but are not
limited to, the sparse optimization of \citep{schaeffer2017learning}, meta-learning of~\citep{feliu2020meta}, as well
as the neural operators of~\citep{
li2021fourier}. A review of relevant 
model identification techniques can be found, e.g., 
in~\citep{karniadakis2021physics}.
Notice that a parallel track to the above one
seeks not to discover the models, but rather
key features thereof such as conserved 
quantities~\citep{teg1,teg2,liu2022machine,PhysRevE.108.L022301} and its potential integrability~\citep{kripy,DEKOSTER2024103273}.

In the present setting, we seek to combine
this important class of dispersive wave models
within optical (and other physical) applications with 
some of the above machine-learning toolboxes.
Our aim is not to discover the full PDE,
or its conservation laws/integrability as in 
some of the above works. Rather, our aim
is to leverage the theoretical understanding
that exists at the level of moment methods~\citep{victor_theory,Victor_NLS_equation}.
Indeed, it is well-known from these works that
upon defining suitable moment quantities,
one can obtain closed form dynamical
systems of a few degrees-of-freedom, often
just two (lending themselves to dynamical systems
analysis) or sometimes involving a few more but
still offering valuable low-dimensional analytical
insights on the evolution of the center of mass,
variance, kurtosis etc. of the relevant distribution.
Indeed, these classes of methods were also
used successfully to other models such as
Fisher-KPP equations, considering also applications
to the dynamics, e.g., of brain tumors~\citep{BELMONTEBEITIA20143267}.

Our approach and presentation will be 
structured hereafter as follows. 
In section 2, we will present a ``refresher''
from a theoretical perspective of the method
of moments, essentially revisiting some basic
results from the work of~\citep{victor_theory,Victor_NLS_equation}.
Then, in section 3, we will give a brief overview
of SINDy type methods and the types of choices
(such as, e.g., of model libraries) that they necessitate. Additionally, we will introduce a data-driven approach for learning coordinate transformations to close moment systems when the initially chosen moments are not closed.
Then, in section 4, we present a palette of numerical
results and their effective moment identification.
Our narrative contains a gradation of examples 
from simpler ones (where, e.g., an analytical
low-dimensional closure of moments may exist)
to gradually more complex ones, where it may exist
after a coordinate transformation and eventually
to cases where a closed system does not exist
at the moment level to the best of our knowledge.
Our aim is to showcase not only the successes, but
also the challenges that the method may encounter
in cases where we do not know of a closure
or when we may not rightfully choose the library
of functions (even when a closure may exist).
We hope that this will provide a more informed/balanced
viewpoint to the reader about what these methods
may (and what they may not) be expected to provide.

\section{Moment Equation Theoretical Background}
To contextualize our perspective, we will focus on the following specific case of the (1+1)-dimensional nonlinear Schr\"{o}dinger (NLS) equation with a harmonic potential, $V(x, t) = \frac{1}{2}x^2$~\citep{siambook,Kivshar2003},
\begin{equation}
\label{eq:nls_harmonic}
    iu_t = -\frac{1}{2}u_{xx} + \frac{1}{2}x^{2}u + g\left(\left|u\right|^{2}, t\right)u,
\end{equation}
where $g\left(|u|^{2},t\right)$ denotes the nonlinearity.
This model is not only of relevance to optics
(where the harmonic potential represents the heterogeneous profile of the refractive index)~\citep{Kivshar2003}, 
but also to
atomic Bose-Einstein condensates, where this
parabolic confinement is a typical byproduct
of magnetic traps~\citep{becbook2,siambook}.
We will consider the initial value problem  of Eq.~\eqref{eq:nls_harmonic} with localized and sufficiently regular initial conditions (ICs) $u_0(x) =u(x, 0)$. 

\noindent \textbf{The method of moments}. Instead of fully characterizing the solution of the Cauchy problem of Eq.~\eqref{eq:nls_harmonic}, the \textit{method of moments \citep{victor_theory}}, seeks to provide qualitative description of the solution behavior by studying the evolution of several integral quantities, i.e., the moments, of the solution $u(x, t)$. This approach enables a reduced-order description of the NLS equation by transforming it into a system of (potentially) closed ordinary differential equations (ODEs). More specifically, according to \citet{victor_theory}, we define for $k = 0, 1, 2, \cdots$ the \textit{moment} quantities of solution $u(x,t)$ as follows,
\begin{align}
    I_{k}(t) &= \int_{\mathbb{R}} x^{k}|u(x,t)|^{2}dx, 
    \label{eq:def_I}\\
    V_{k}(t) &= 2^{k-1}i\int_{\mathbb{R}}  x^{k}\left(u(x,t)\frac{\partial \bar{u}(x,t)}{\partial x} - \bar{u}(x,t)\frac{\partial u(x,t)}{\partial x}\right)dx, 
    \label{eq:def_V}\\
    K(t) &= \frac{1}{2}\int_{\mathbb{R}}  \left|\frac{\partial u(x,t)}{\partial x}\right|^{2} dx, 
    \label{eq:def_K}\\
    J(t) &= \int_{\mathbb{R}}  G\left(\rho(x, t), t\right)dx = \int_{\mathbb{R}}  G\left(\left|u(x, t)\right|^2, t\right)dx,
    \label{eq:def_J}
\end{align}
where $\bar{u}$ is the complex conjugate of $u$, $\rho(x, t) = \left|u(x, t)\right|^{2}$, and $G=G(\rho, t)$ is a function such that $\frac{\partial G}{\partial \rho}(\rho, t) = g(\rho, t)$. Moments \eqref{eq:def_I}, \eqref{eq:def_V}, \eqref{eq:def_K}, and \eqref{eq:def_J} of the solution $u(x,t)$ have intuitive physical meanings; for example, the first moment $I_1(t)$ is associated with the center of mass as described by the (unnormalized) probability density $\rho = |u|^2$. 
Higher moments $I_k$ are also associated with 
this density distribution (i.e., its variance etc.).
The $V_k$ quantities are the respective ones associated
with the momentum density (which is the quantity
in the corresponding parenthesis in the right hand
side of the $V_k$ definition. $K$ stems from  
(thinking quantum-mechanically) the kinetic part of
the Schr{\"o}dinger problem energy, while $J$
represents the nonlinear part of the corresponding
energy.
We assume that the IC $u_0(x)$ is regular enough to ensure that all moments are well-defined for $t\ge 0$.

The method of moments aims to extract qualitative information  about the solution $u(x,t)$ of the PDE~\eqref{eq:nls_harmonic} by deriving a \textit{closed} set of evolution ODEs for the moments of $u(x,t)$. Depending on the nonlinearity $g(\rho, t)$, these ODEs can sometimes be determined \textit{analytically}
as is shown in the work of~\citep{victor_theory,Victor_NLS_equation}. Below, we provide a few examples.

\begin{example}
\label{ex:ex_1}
The moments $I_1$ and $V_0$ satisfy
    \begin{align}
    \label{eq:first_ME}
    \left\{
        \begin{aligned}
            \frac{dI_1}{dt} &= V_0,\\
            \frac{dV_0}{dt} &= -I_1.
        \end{aligned}\right.
    \end{align}
    This indicates that the evolution of the center of mass $I_1$ behaves as a harmonic oscillator, independent of the nonlinearity $g(\rho, t)$.
    More generally, for a parabolic confinement 
    of frequency $\Omega$, this would be reflected
    in the associated frequency of moment oscillations;
    this is the so-called dipolar motion~\citep{becbook2,siambook}.
\end{example}

\begin{example}
\label{ex:ex_2}
If $g(\rho, t)\equiv 0$, i.e., for the linear case of \eqref{eq:nls_harmonic}, the set of moments $I_2, V_1, K$ are closed under
    \begin{align}
    \label{eq:second_ME}
    \left\{
    \begin{aligned}
        \frac{dI_2}{dt} &= V_1,\\
        \frac{dV_1}{dt} &= 4K - 2I_2,\\
        \frac{dK}{dt} &= -\frac{1}{2}V_1.
    \end{aligned}\right.
    \end{align}
\end{example}

\begin{example}
\label{ex:ex_3}
Assume the nonlinearity $g(\rho, t) = g(\rho)$ is time-independent and given by $g(\rho) = g_0\rho^{2}$, where $g_0\in \R$ is a constant. Although the evolution of the moments $I_2$, $V_1$, $K$ and $J$ is not closed, it becomes closed under the coordinate transformation $E = K+J$, i.e.,
    \begin{align}
    \label{eq:third_ME}
    \left\{
    \begin{aligned}
        \frac{dI_2}{dt} &= V_1,\\
        \frac{dV_1}{dt} &= 4E - 2I_2,\\
        \frac{dE}{dt} &= -\frac{1}{2}V_1,
    \end{aligned}\right.
    \end{align}
\end{example}

While the examples and conditions under which the method of moments leads to closed equations are well-known, deriving such analytical closure systems requires  knowledge of the underlying PDE system, Eq.~\eqref{eq:nls_harmonic} and
detailed calculations therewith. This work explores \textit{data-driven} methods for obtaining analytical or approximate moment closure systems based on empirical \textit{observations} or \textit{simulations} of the NLS equation, rather than relying solely on such analytical
understanding and derivations. Importantly,
the reconstruction of these ODE models
can, in principle, take place even for settings
where the underlying PDE model is unavailable/has
not been specified. Given ``experimental'' data
for the field, one may aspire to utilize the toolboxes
presented below in order to obtain these effective,
reduced dynamical equations.

For systems with existing analytical closures of the moment equations, such as Examples~\ref{ex:ex_1}-\ref{ex:ex_3}, our method seeks to rediscover the moment evolution equations and potentially the necessary coordinate transformations, such as $E = K+J$ in Example~\ref{ex:ex_3}, in a model-agnostic and data-driven manner. For systems lacking analytical closed moment equations, we seek to derive approximate moment closure equations, providing a principled and reduced-order description of the original PDE, capable of predicting the future evolution of the system. The relevant details will be explained in the following section.

\section{Data-Driven Methods}
\label{sec:methods}
We present two computational methodologies for finding analytical or approximate closures for moment equations.

\subsection{Sparse Identification of Nonlinear Dynamics}

Our first method leverages Sparse Identification of Nonlinear Dynamics (SINDy) by \citet{Kutz_SINDy}, a data-driven approach for discovering governing ODEs from simulated or observational data. Consider a nonlinear ODE system of the form:
\begin{align}
\label{eq:ode}
    \frac{d\bfx}{dt}(t) = \bff(\bfx (t)),
\end{align}
where $\bfx = (x_1, \cdots, x_n)^\top :[0, \infty)\to \R^n$ represents the state evolution over time, governed by the dynamic constraint $\bff:\R^n\to\R^n$. SINDy aims to identify the unknown dynamics, \(\bff(\bfx)\) from a time series of \(\bfx\).

The key assumption is that \(\bff(\bfx)\) has a ``simple'' form and can be expressed or approximated as a linear combination of only a few terms from a suitably chosen library, $\mathbf{\Theta}(\bfx) = [\theta_1(\bfx), \cdots, \theta_p(\bfx)]$. For example, a monomial library of degree  up to two, $\mathbf{\Theta}_{\deg \le 2}(\bfx)$, is:
\begin{align}
\label{eq:deg_2_lib}
    \mathbf{\Theta}_{\deg \le 2}(\bfx) = \big[\mathbf{\Theta}_{\deg = 1}(\bfx), \mathbf{\Theta}_{\deg = 2}(\bfx)\big]    
    & = \big[\underbrace{x_1, x_2, \cdots, x_n}_{\mathbf{\Theta}_{\deg = 1}(\bfx)}, \underbrace{x_1^2, x_1x_2, \cdots, x_1x_n, x_2^2, \cdots, x_n^2}_{\mathbf{\Theta}_{\deg = 2}(\bfx)}\big]\\
    & = \left[\theta_1(\bfx), \cdots, \theta_p(\bfx)\right],
\end{align}
with $p = n + {n \choose 2}$.
In particular, e.g., the right-hand sides of Eqs.~\eqref{eq:first_ME}-\eqref{eq:third_ME} can all be written as sparse linear combinations of terms from $\mathbf{\Theta}_{\deg \le 2}(\bfx)$. Given a dictionary $\mathbf{\Theta}(\bfx) = [\theta_1(\bfx), \cdots, \theta_p(\bfx)]$ of $p$ atoms---$p$ is typically larger than $n$---the sparsity assumption implies the existence of a sparse matrix $\mathbf{\Xi} = (\bm{\xi}_1, \cdots, \bm{\xi}_n)\in\R^{p\times n}$ such that
\begin{align}
\label{eq:sparse_relation_function}
    \bff(\bfx)^\top
    =
    \begin{bmatrix}
        f_1(\bfx), &\cdots, &f_n(\bfx)
    \end{bmatrix}
    \approx \mathbf{\Theta}(\bfx)\cdot \mathbf{\Xi}
    & = \left[\theta_1(\bfx), \cdots, \theta_p(\bfx)\right]
    \cdot
    \begin{bmatrix}
        | & | & \cdots &|\\
        \bm{\xi}_1 & \bm{\xi}_2 & \cdots & \bm{\xi}_n\\
        | & | & \cdots &|        
    \end{bmatrix}
\end{align}
where  each sparse column $\bm{\xi}_j\in\R^p$ indicates which nonlinear functions among the library $\mathbf{\Theta}(\bfx)=[\theta_1(\bfx), \cdots, \theta_p(\bfx)]$ are used to parsimoniously represent $f_j(\bfx)$.

To determine this sparse $\mathbf{
\Xi}$, SINDy employs sparse regressions on the data. Specifically, given a time series \(\{\bfx(t_1), \cdots, \bfx(t_N)\}\subset \R^n\) of the state $\bfx(t)$ at times $t_1, \cdots, t_N$---$N$ is generally much larger than $p$, the library size---one can assemble the \textit{state matrix} $\mathbf{X}\in \R^{N\times n}$ and the \textit{derivative matrix} $\dot{\mathbf{X}}\in \R^{N\times n}$:
\begin{align}
\label{eq:def_state_matrix}
    \mathbf{X}
    =
    \begin{bmatrix}
        | & | & & |\\
        \bfX_1 & \bfX_2 &\cdots &\bfX_n\\
        | & | & & |
    \end{bmatrix}    
    \coloneqq
    \begin{bmatrix}
        x_1(t_1) & x_2(t_1) & \cdots & x_n(t_1)\\
        x_1(t_2) & x_2(t_2) & \cdots & x_n(t_2)\\
        \vdots & \vdots & \ddots & \vdots\\
        x_1(t_N) & x_2(t_N) & \cdots & x_n(t_N)
    \end{bmatrix}
    \in \R^{N\times n},
\end{align}
\begin{align}
\label{eq:def_derivative_matrix}
    \frac{d}{dt}\mathbf{X}
    =\dot{\mathbf{X}}
    =
    \begin{bmatrix}
        | & | & & |\\
        \dot{\bfX}_1 & \dot{\bfX}_2 &\cdots &\dot{\bfX}_n\\
        | & | & & |
    \end{bmatrix}    
    \coloneqq
    \begin{bmatrix}
        \dot{x}_1(t_1) & \dot{x}_2(t_1) & \cdots & \dot{x}_n(t_1)\\
        \dot{x}_1(t_2) & \dot{x}_2(t_2) & \cdots & \dot{x}_n(t_2)\\
        \vdots & \vdots & \ddots & \vdots\\
        \dot{x}_1(t_N) & \dot{x}_2(t_N) & \cdots & \dot{x}_n(t_N)
    \end{bmatrix}
    \in \R^{N\times n},
\end{align}
where $\dot{\bfX}$ can be estimated by, e.g., finite differences on $\bfX$. Define the \textit{library matrix} $\mathbf{\Theta}(\bfX)\in\R^{N\times p}$ as
\begin{align}
\label{eq:library_matrix}
    \mathbf{\Theta}(\bfX)
    =
    \begin{bmatrix}
        | & | & & |\\
        \theta_1(\bfX) & \theta_2(\bfX) &\cdots &\theta_p(\bfX)\\
        | & | & & |
    \end{bmatrix}
    & \coloneqq
    \begin{bmatrix}
        \theta_1(\bfx(t_1)) & \theta_2(\bfx(t_1)) & \cdots & \theta_p(\bfx(t_1))\\
        \theta_1(\bfx(t_2)) & \theta_2(\bfx(t_2)) & \cdots & \theta_p(\bfx(t_2))\\
        \vdots & \vdots & \ddots & \vdots\\
        \theta_1(\bfx(t_N)) & \theta_2(\bfx(t_N)) & \cdots & \theta_p(\bfx(t_N))
    \end{bmatrix}.
\end{align}
Evaluating Eq.~\eqref{eq:ode} and Eq.~\eqref{eq:sparse_relation_function} at all times $t=t_1, \cdots, t_N$ yields
\begin{align}
\label{eq:sparse_relation_matrix}
    \begin{bmatrix}
        | & | & & |\\
        \dot{\bfX}_1 & \dot{\bfX}_2 &\cdots &\dot{\bfX}_n\\
        | & | & & |
    \end{bmatrix}
    =
    \dot{\bfX}
    \approx
    \mathbf{\Theta}(\bfX)\bm{\Xi}
    =
    \begin{bmatrix}
        | &  & |\\
        \theta_1(\bfX) & \cdots &\theta_p(\bfX)\\
        | &  & |
    \end{bmatrix}
    \begin{bmatrix}
        | & | & \cdots &|\\
        \bm{\xi}_1 & \bm{\xi}_2 & \cdots & \bm{\xi}_n\\
        | & | & \cdots &|        
    \end{bmatrix}
\end{align}
Sparse regression techniques, such as LASSO \citep{tibshirani1996regression} or sequential thresholded least-squares \citep{Kutz_SINDy}, can then solve the overdetermined system \eqref{eq:sparse_relation_matrix} for the sparse $\bm{\Xi}$. This provides an approximation to the governing equation as in Eq~(\ref{eq:sparse_relation_function}).

While SINDy can be directly applied to Examples~\eqref{ex:ex_1}-\eqref{ex:ex_3} to discover closed moment systems based on simulated PDE data [Eq.~\eqref{eq:nls_harmonic}],  our specific interest lies in the following:
\begin{enumerate}
    \item Given time series data of moments with a known closed moment system (e.g., Examples \eqref{ex:ex_1} and \eqref{ex:ex_2}), can we robustly discover the governing dynamics using a correct but potentially oversized dictionary $\mathbf{\Theta}(\bfx)$, such as polynomials up to a high degree?
    
    \item If a system is not closed for a chosen set of moments, but closure exists after a proper coordinate transformation (e.g., selecting $\bfx=[I_2, V_1, K, J]^\top$ in Ex.~\ref{ex:ex_3}), what insights can we gain by directly applying SINDy to time series data from this ``incorrect'' set of moments?
    
    \item For more general systems where analytical closure does not exist, can data-driven methods provide a ``good enough" numerical approximation to predict the future evolution of the moment system, effectively serving as a principled reduced-order description of the underlying PDE?
\end{enumerate}
These questions will be addressed in Section~\ref{sec:experiments}. Of particular interest is the second point, where we demonstrate that, in certain cases, we can gain insight into the appropriate transformation to close the system, even if the moment system is not closed under the originally selected 
variables. In the following section, we discuss a more principled strategy to discover such transformation in a data-driven fashion.

\subsection{Data-driven discovery of coordinate transformations for moment system closure}
\label{sec:stiefel_opt}
To illustrate the idea, we focus on Example~\eqref{ex:ex_3}, where the initially selected moments are $\bfx=[I_2, V_1, K, J]^\top$, and an analytical closure exists \textit{only after} a coordinate transformation, $\bfy = \mathbf{A}^\top \bfx$,
\begin{align}
    \bfy = 
    \begin{bmatrix}
        I_2\\
        V_1\\
        K+J
    \end{bmatrix}
    =
    \begin{bmatrix}
        1 & 0 & 0 & 0\\
        0 & 1 & 0 & 0\\
        0 & 0 & 1 & 1
    \end{bmatrix}
    \begin{bmatrix}
        I_2\\
        V_1\\
        K\\
        J
    \end{bmatrix}
    =
    \mathbf{A}^\top \bfx,    
\end{align}
\begin{remark}
\label{rmk:rmk_1}
    Note that the transformation matrix $\mathbf{A} \in \R^{4 \times 3}$ is not unique. For any full-rank matrix $\mathbf{P} \in \R^{3 \times 3}$, the moment system remains closed under the transformation $\widetilde{\bfy} = \widetilde{\mathbf{A}}^\top \bfx$, with $\widetilde{\mathbf{A}} = \mathbf{A}\mathbf{P}$.
\end{remark}
Assume we aim to discover $\mathbf{A}$ purely from simulated PDE data. We propose the following strategy: Let $\bfX\in \R^{N\times 4}$ be the state matrix for the original coordinate $\bfx=[I_2, V_1, K, J]^\top$, sampled at $t_1, \cdots, t_N$, as defined by Eq.~\ref{eq:def_state_matrix}. Let $\bfy = \mathbf{A}^\top \bfx$ be the new coordinate, and the associated new state matrix $\bfY\in\R^{N\times 3}$ becomes
\begin{align}
    \bfY = 
    \begin{bmatrix}
        \text{---}~\bfy(t_1)^\top~\text{---}\\
        \text{---}~\bfy(t_2)^\top~\text{---}\\
        \vdots\\
        \text{---}~\bfy(t_N)^\top~\text{---}
    \end{bmatrix}
    =
    \begin{bmatrix}
        \text{---}~\bfx(t_1)^\top~\text{---}\\
        \text{---}~\bfx(t_2)^\top~\text{---}\\
        \vdots\\
        \text{---}~\bfx(t_N)^\top~\text{---}
    \end{bmatrix}
    \mathbf{A}
    =
    \bfX \mathbf{A}
\end{align}
We can then solve for $\mathbf{A}$ through the following optimization
\begin{align}
\label{eq:optimization_inner_outer}
    & \min_{\mathbf{A}\in \R^{4\times 3}}\min_{\mathbf{\Xi} \in \R^{p\times 3}} \left\|\frac{d}{dt}\left(\mathbf{XA}\right)- \mathbf{\Theta}\left(\mathbf{XA}\right)\cdot \mathbf{\Xi} \right\|^2_F + \mu \|\mathbf{\Xi}\|_1, \\
    & \text{ s.t.~}   \mathbf{A}^\top \mathbf{A} = \mathbf{I}_{3\times 3}  
\end{align}
where $\frac{d}{dt}\left(\mathbf{XA}\right)$ is the derivative matrix, Eq.~\ref{eq:def_derivative_matrix}, associated with the new state matrix $\bfY = \mathbf{XA}$, $\|\cdot\|_F$ is the Frobenius norm, $\mu\ge0$ is a non-negative weight, and $\mathbf{\Theta}(\mathbf{XA})=\mathbf{\Theta}(\mathbf{Y})\in\R^{N\times p}$ is the library matrix, Eq.~\ref{eq:library_matrix}, of a chosen library $\mathbf{\Theta}(\bfy)$ on the new coordinate $\bfy = \mathbf{A}^\top \bfx$.

The idea behind Eq.~\eqref{eq:optimization_inner_outer} is very simple: we search for the transformation matrix $\mathbf{A}\in\R^{4\times 3}$ such that the dynamics under the new coordinate $\bfy = \mathbf{A}^\top \bfx$ can be parsimoniously represented from the library $\mathbf{\Theta}(\bfy)$. The weight $\mu\ge 0$ controls the sparsity-promoting $L_1$-regularization, and the constraint $\mathbf{A}^\top\mathbf{A}=\mathbf{I}_{3\times 3}$  prevents the trivial solution $\mathbf{A}=\mathbf{0}$. The set of $\mathbf{A}$ satisfying the constraint $\mathbf{A}^\top\mathbf{A}=\mathbf{I}_{3\times 3}$ is called the \textit{Stiefel manifold} \citep{edelman1998geometry,absil2008optimization}.
\begin{remark}
\label{rmk:rmk_2}
    When the weight $\mu= 0$, even with the Stiefel manifold constraint $\mathbf{A}^\top \mathbf{A}=\mathbf{I}_{3\times 3}$, the solution to Eq.~\ref{eq:optimization_inner_outer} is not unique. Just like Remark~\ref{rmk:rmk_1}, if $(\mathbf{A}^*, \mathbf{\Xi}^*)$ is a solution, then for any orthogonal matrix $\mathbf{O}\in \text{O}(3)$, the pair $(\widetilde{\mathbf{A}}^*, \widetilde{\mathbf{\Xi}}^*)$ is also a solution, where
    \begin{align}
    \widetilde{\mathbf{A}}^* = \mathbf{A}^*\mathbf{O}, \quad \widetilde{\mathbf{\Xi}}^* = \mathbf{O}^\top \mathbf{\Xi}^*\mathbf{O}
    \end{align}
\end{remark}

To solve Eq.~\eqref{eq:optimization_inner_outer}, we use alternating optimization, iteratively optimizing $\mathbf{A}$ and $\mathbf{\Xi}$ while keeping the other fixed. Specifically, when $\mathbf{A}$ is fixed, solving for $\mathbf{\Xi}$ reduces to a LASSO problem. Conversely, when $\mathbf{\Xi}$ is fixed, the problem becomes a Stiefel manifold optimization with a smooth objective function, which can be efficiently solved using methods such as those presented in \citep{10.1007/978-3-030-33749-0_20}
\citep{doi:10.1080/10556788.2020.1852236}
\citep{Liu2021API}. To ensure the algorithm remains unbiased, we implement an annealing strategy for the hyperparameter \(\mu\). Initially, \(\mu\) is kept constant for \texttt{Iter\_scheduled} iterations. Following this period, \(\mu\) is reduced by half every \(\alpha\) iterations.
Refer to Algorithm\ref{alg:inner_outer_optimization_new} for the pseudocode to solve problem~\eqref{eq:optimization_inner_outer}.

 

\begin{algorithm}[h]
\caption{Data-driven discovery of coordinate transformation}
\label{alg:inner_outer_optimization_new}
\KwInput{\texttt{maxIter}: Total number of iterations.\\
\hspace{4.5em}\texttt{Iter\_scheduled}: Number of initial iterations during which \(\mu\) remains constant.\\
\hspace{4.5em}\texttt{$\alpha$}: Number of iterations after which \(\mu\) is halved.\\
\hspace{4.5em}Random initializations: $\mathbf{A}^{(0)}\in\R^{4\times 3}$ on the Stiefel manifold and $\mathbf{\Xi}^{(0)}\in\R^{p\times 3}$.}
\KwOutput{\hspace{.8em}$\mathbf{A}_{\text{out}}$, $\mathbf{\Xi}_{\text{out}}$}
 
\For{$k$ in $[1, 2, \cdots, \tt{maxIter}]$}{
    \If{$k > \texttt{Iter\_scheduled}$ and $k$ is a multiple of $\alpha$}
    {
        $\mu \gets \mu / 2$\;
    }
  $\mathbf{\Xi}^{(k)} \leftarrow \arg\min_{\mathbf{\Xi}} \left\|\frac{d}{dt}\left(\mathbf{XA}^{(k-1)}\right)- \mathbf{\Theta}\left(\mathbf{XA}^{(k-1)}\right)\cdot \mathbf{\Xi} \right\|^2_F + \mu \|\mathbf{\Xi}\|_1$, solved by LASSO.\\
  $\mathbf{A}^{(k)} \leftarrow \arg\min_{\mathbf{A}^\top \mathbf{A} = \mathbf{I}} \left\|\frac{d}{dt}\left(\mathbf{XA}\right)- \mathbf{\Theta}\left(\mathbf{XA}\right)\cdot \mathbf{\Xi}^{(k-1)} \right\|^2_F$, solved by Stiefel optimization.
}
$\mathbf{A}_{\text{out}}\leftarrow \mathbf{A}^{(\tt{maxIter})}$\\
$\mathbf{\Xi}_{\text{out}}\leftarrow \mathbf{\Xi}^{(\tt{maxIter})}$
\end{algorithm}

\section{Numerical Results}
\label{sec:experiments}

In this section, we present numerical results on data-driven closure of moment systems (Examples \eqref{ex:ex_1}-\eqref{ex:ex_3}) using the methodologies described in Section~\ref{sec:methods}. To obtain the data, we first numerically solve the PDE~\eqref{eq:nls_harmonic} with periodic boundary conditions and various ICs using an extended 4th-order Runge-Kutta method, the exponential integrator (ETDRK4) (see \citep{4thordeRK4} for a detailed explanation). The time series of the moments is
subsequently extracted by numerical spatial integration according to Eqs.~\eqref{eq:def_I}-\eqref{eq:def_J}, with spatial derivatives computed using pseudo-spectral Fourier methods.  
We note once again that should the numerical data
be replaced by ``experimental'' ones from a given
physical process, the procedure can still be applied.
Unless otherwise noted, the time series are evaluated at $N=16,000$ uniformly spaced times $t = t_1, \cdots, t_N$, where $t_i = i \Delta t$ and $\Delta t = 0.0025$.

\subsection{Examples with analytical moment closure}
We first examine Examples \eqref{ex:ex_1} and \eqref{ex:ex_2}, where analytical (linear) closure exists for the chosen moments $\bfx = [I_1, V_0]$ (Example~\eqref{ex:ex_1}) and $\bfx = [I_2, V_1, K]$ (Example~\eqref{ex:ex_2}).

\subsubsection{Example~\ref{ex:ex_1}}
\label{sec:ex_1}
Let the selected moments be $\bfx = [I_1, V_0]$. We construct the data matrices $\bfX^{(0)} = [\mathbf{I}_1^{(0)}, \mathbf{V}_0^{(0)}] \in \R^{N \times 2}$ and $\bfX^{(1)} = [\mathbf{I}_1^{(1)}, \mathbf{V}_0^{(1)}]\in \R^{N \times 2}$ from numerically solving the PDE~\eqref{eq:nls_harmonic} with the ICs:
\begin{align}
    u^{(1)}(x,0) &= \pi^{-1/4}\exp\left(-\frac{1}{2}\left(x-5\right)^{2}\right),
    \label{eq:ex_1_IC_1}\\
    u^{(2)}(x,0) &= \frac{1}{2}\sech^{2}\left(x-5\right).
    \label{eq:ex_1_IC_2}
\end{align}
We apply SINDy to this system using monomial libraries of degree up to $n \in \mathbb{N}$, $\mathbf{\Theta}_{\deg\le n}(\bfx)$, as defined in Eq.~\eqref{eq:deg_2_lib}.
We find that as long as $n \le 2$, SINDy applied to either data matrix $\bfX^{(0)}$ or $\bfX^{(1)}$ discovers the dynamics nearly perfectly. However, as we gradually expand the library $\mathbf{\Theta}_{\deg\le n}(\bfx)$ by increasing $n$, SINDy applied to either $\bfX^{(0)}$ or $\bfX^{(1)}$ \textit{individually} typically produces erroneous dynamics, regardless of how carefully the sparsity-promoting parameter is tuned. A typical negative result from SINDy applied to $\bfX^{(0)}$ with $n=3$ is presented in the appendix. This outcome is expected, as increasing the library size makes the problem more ill-posed, leading SINDy to overfit the data and produce incorrect dynamics.

To address the overfitting issue, we can enlarge the dataset by concatenating the data matrices $\bfX^{(0)}$ and $\bfX^{(1)}$ vertically, forming $\bfX = [\bfX^{(0)\top}, \bfX^{(1)\top}]^\top \in \R^{2N \times 2}$. When applying SINDy to this new data matrix $\bfX$ (considering boundary issues when taking finite differences), SINDy can now discover the correct dynamics that match Eq.~\eqref{eq:first_ME}, even with a much larger library $\mathbf{\Theta}_{\deg\le n}(\bfx)$ for $n$ up to 16. For example, when $n = 16$, the output ODE from SINDy reads
\begin{equation}\label{eq: SINDy output for n = 16}
\left\{
    \begin{aligned}
        \frac{dI_1}{dt} &= 1.000V_0,\\
        \frac{dV_0}{dt} &= -1.000I_1,
    \end{aligned}\right.
\end{equation}
where the coefficients are rounded to three decimal places.
This suggests the potential usefulness of concatenating
different time series, especially in cases where
one may not be familiar with the order of the relevant closure.

\subsubsection{Example~\ref{ex:ex_2}}
For Example \ref{ex:ex_2} with the selected moments $\bfx = [I_2, V_1, K]$, our findings are similar to those in Section~\ref{sec:ex_1}. When applying SINDy to a data matrix from a single IC, SINDy discovers erroneous dynamics for the quadratic dictionary $\mathbf{\Theta}_{\deg\le 2}$. However, using larger data matrices from multiple ICs, SINDy can once again accurately identify the correct dynamics even with the larger dictionaries. Representative negative and positive results are presented in the appendix, similar to the previous example.

\subsection{Examples where closure exists after coordinate transformations}
We now turn to Example \eqref{ex:ex_3}, where the selected moments are $\bfx = [I_2, V_1, K, J]$, and the moment system only closes after a coordinate transformation. Specifically, we consider the nonlinearity $g(\rho, t) = \rho^2$ in Eq.~\eqref{eq:nls_harmonic}, which satisfies the condition in Example \eqref{ex:ex_3}. We collect the moment time series data by numerically solving the PDE with the following four distinct ICs, Eqs.~\eqref{eq:ex_3_IC_1}, \eqref{eq:ex_3_IC_2}, \eqref{eq:ex_3_IC_3} and \eqref{eq:ex_3_IC_4}:
\begin{align}
    u^{(1)}(x,0) &= \pi^{-1/4}\exp\left(-\frac{1}{2}\left(x-5\right)^{2}\right),
    \label{eq:ex_3_IC_1}\\
    u^{(2)}(x,0) &= 1.88\exp\left(-\frac{1}{2}\left(x-5\right)^{2}\right),
    \label{eq:ex_3_IC_2}\\
    u^{(3)}(x,0) &= 1.88\left(\exp\left(-\frac{1}{2}\left(x-5\right)^{2}\right) + \exp\left(-\left(x-2\right)^{2}\right)\right),
    \label{eq:ex_3_IC_3}\\
    u^{(4)}(x,0) &=  1.88\left(\cos(2x) + \sin(2x)\right)\exp(-x^{2}).
    \label{eq:ex_3_IC_4}
\end{align}
We consider the following questions:
\begin{itemize}
    \item What insights can we gain by directly applying SINDy to this system with the selected moments \(\bfx = [I_2, V_1, K, J]\), where a closure does not exist?
    \item Can the method described in Section~\ref{sec:stiefel_opt} correctly identify the coordinate transformation that closes the moment system?
\end{itemize}
We note that \textit{normalized} moment time series are used as input for SINDy, as a way of incorporating feature scaling.
This is an important aspect that ensures that
all the relevant quantities are considered
on ``equal footing'', when the sparse regression
step takes place.

\subsubsection{SINDy with linear library $\mathbf{\Theta}_{\deg = 1}(\bfx)$}
\label{sec:ex_3_sindy_lib_1}
\begin{figure}[t]
\begin{center}
\includegraphics[width=0.85\linewidth]{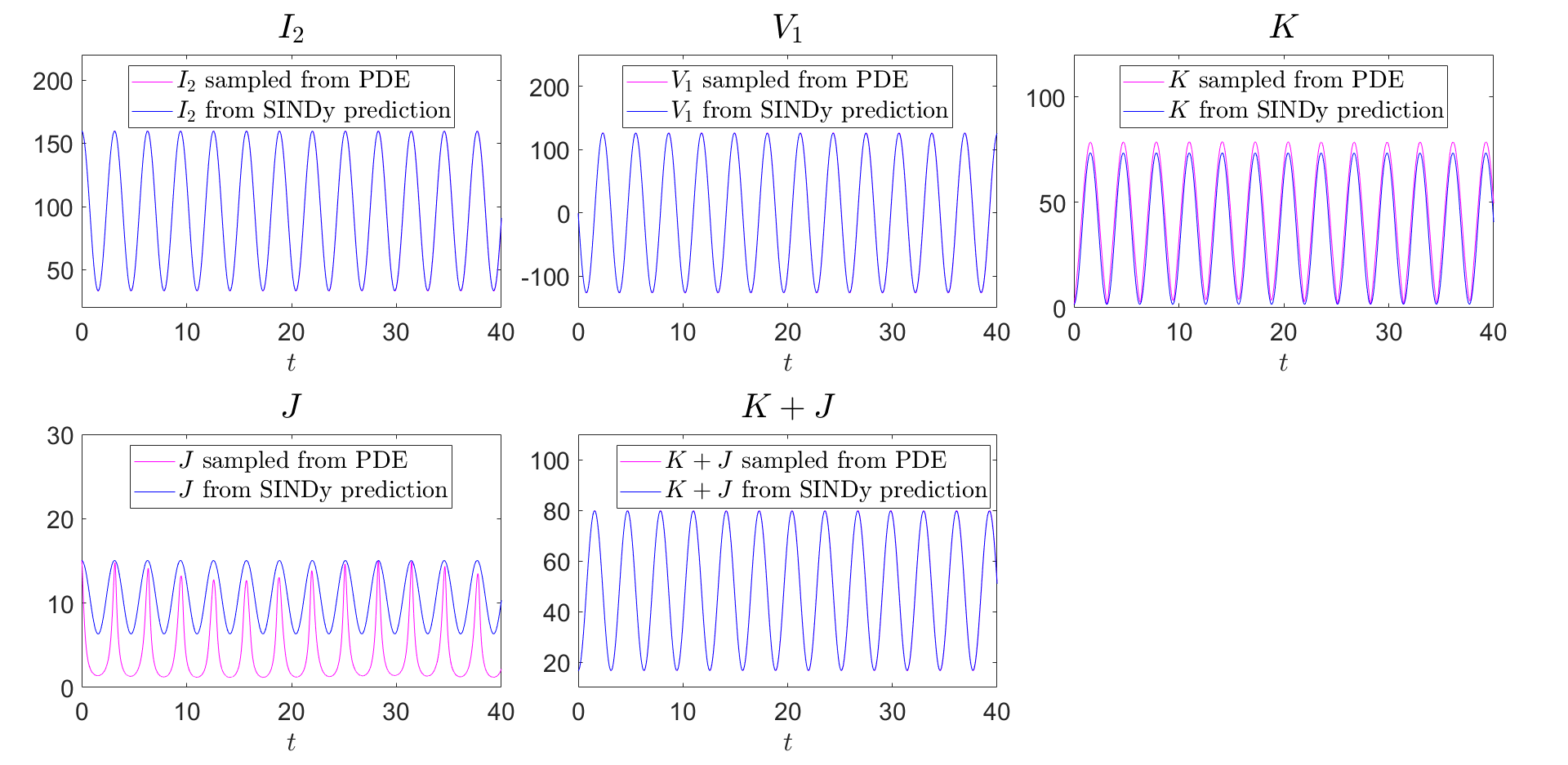}
\end{center}
\caption{Comparison of the ground-truth and SINDy-predicted time evolutions of \([I_2, V_1, K, J, K+J]\). SINDy is trained only on the selected moments \(\bfx = [I_2, V_1, K, J]\), where a closure does not exist, using a linear library \(\mathbf{\Theta}_{\deg=1}(\bfx)\). Interestingly, SINDy 
can ``afford''
to sacrifice accurate prediction for \(K\) and \(J\) \textit{individually}, provided that it captures the correct dynamics for \(E=K+J\), suggesting in this way
the proper coordinate transformation under which a closure does exist.}\label{fig:ex_3_sindy_lib_1}
\end{figure}

We begin by applying SINDy with a linear library $\mathbf{\Theta}_{\deg = 1}(\bfx)$ to the moment time series data from IC~\eqref{eq:ex_3_IC_2}. The resulting equations are:
\begin{align}
\label{eq:ex_3_sindy_raw_lib_1}
\left\{
\begin{aligned}
    \frac{dI_2}{dt} &= 1.000V_1,\\
    \frac{dV_1}{dt} &= -2.000I_2 + 4.000K + 3.998J,\\
    \frac{dK}{dt} &= -0.569V_1,\\
    \frac{dJ}{dt} &= 0.069V_1.
\end{aligned}\right.
\end{align}
The coefficients are rounded to three decimal places. The first four panels of Figure~\ref{fig:ex_3_sindy_lib_1} compare the ``ground-truth'' time-evolution data of $[I_2, V_1, K, J]$ obtained from PDE integration with those from integrating the SINDy-predicted ODEs~\eqref{eq:ex_3_sindy_raw_lib_1}. While the SINDy-predicted time evolution of $[I_2, V_1]$ closely matches the ground truth, there is a {\it significant discrepancy} for the moments $[K, J]$ in Figure~\ref{fig:ex_3_sindy_lib_1}. Accordingly, the SINDy-predicted dynamics is not accurate for the original coordinates $\bfx=[I_2, V_1, K, J]$.

Nevertheless, interestingly, if we add the ODE for $K$ with that for $J$ in Eq.~\eqref{eq:ex_3_sindy_raw_lib_1}, which corresponds to the \textit{correct} coordinate transformation for closure, we obtain:
\begin{align}
 \label{eq:ex_3_sindy_transformed_lib_1}
\left\{
\begin{aligned}
    \frac{dI_2}{dt} &= 1.000V_1,\\
    \frac{dV_1}{dt} &= -2.000I_2 + 4.000K + 3.998J,\\
    \frac{d\left(K+J\right)}{dt} &= -0.500V_1.
\end{aligned}\right.
\end{align}
Eq.~\eqref{eq:ex_3_sindy_transformed_lib_1} closely matches the ground-truth moment system \eqref{eq:third_ME}. The last panel of Figure~\ref{fig:ex_3_sindy_lib_1} shows that the predicted time evolution of \(K+J\) aligns perfectly with the ground truth, even though SINDy does not accurately recover those of \(K\) and \(J\) \textit{individually}. This finding is intriguing because, even when SINDy is applied to the original coordinates \([I_2, V_1, K, J]\) without closure, it ``strategically chooses'' to sacrifice accuracy for \(K\) and \(J\) but indirectly captures the correct dynamics when  considering the proper coordinate transformation $E = K+J$.
Moreover, a similar pattern is consistently observed when applying SINDy to the moments simulated from the other three ICs. The results are shown in the appendix.

\subsubsection{SINDy with quadratic library $\mathbf{\Theta}_{\deg \le 2}(\bfx)$}
\label{sec:blowup}
Next, we investigate the performance of SINDy with an expanded quadratic library $\mathbf{\Theta}_{\deg \le 2}(\bfx)$ applied to the moment time series with the same IC \eqref{eq:ex_3_IC_2}. The predicted ODEs now become:
\begin{align}
\label{eq:ex_3_sindy_raw_lib_2}
\left\{
\begin{aligned}
    \frac{dI_2}{dt} &= 1.000V_1,\\
    \frac{dV_1}{dt} &= -2.000I_2+4.000K+3.998J,\\
    \frac{dK}{dt} &= -0.174V_1-0.003V_1K - 0.083V_1J,\\
    \frac{dJ}{dt} &= -0.002I_2V_1 + 0.082V_1J.
\end{aligned}\right.
\end{align}
As before, if we add the predicted dynamics of $K$ and $J$ in Eq.~\eqref{eq:ex_3_sindy_raw_lib_2}, we obtain:
\begin{align}
\label{eq:ex_3_sindy_transformed_lib_2}
\left\{
\begin{aligned}
    \frac{dI_2}{dt} &= 1.000V_1,\\
    \frac{dV_1}{dt} &= -2.000I_2+4.000K+3.998J,\\
    \frac{d\left(K+J\right)}{dt} &= -0.174V_1 - 0.003V_1K - 0.002I_2V_1 - 0.001V_1J.
\end{aligned}\right.
\end{align}
Unlike the previous case with \(\mathbf{\Theta}_{\deg=1}(\bfx)\), the SINDy-predicted governing equations~\eqref{eq:ex_3_sindy_transformed_lib_2} using the expanded library \(\mathbf{\Theta}_{\deg\le 2}(\bfx)\) after the (theoretically motivated) coordinate transformation $E = K+J$ still fail to match the ground truth ODE \eqref{eq:third_ME} and remain unclosed (i.e., the third equation in \eqref{eq:ex_3_sindy_transformed_lib_2} cannot be written using only $I_2$, $V_1$, and $E=K+J$). Specifically, the coefficient of $V_1$ in Eq.~\eqref{eq:ex_3_sindy_transformed_lib_2} is  $-0.174$, deviating substantially from that of $-0.5$ in the ground truth.  Although the coefficients for the additional terms ($V_1K$, $I_2V_1$, and $V_1J$) are relatively small, their presence hinders the accurate recovery of the correct coefficient for $V_1$. 

\begin{figure}[t]
\begin{center}
\includegraphics[width=0.85\linewidth]{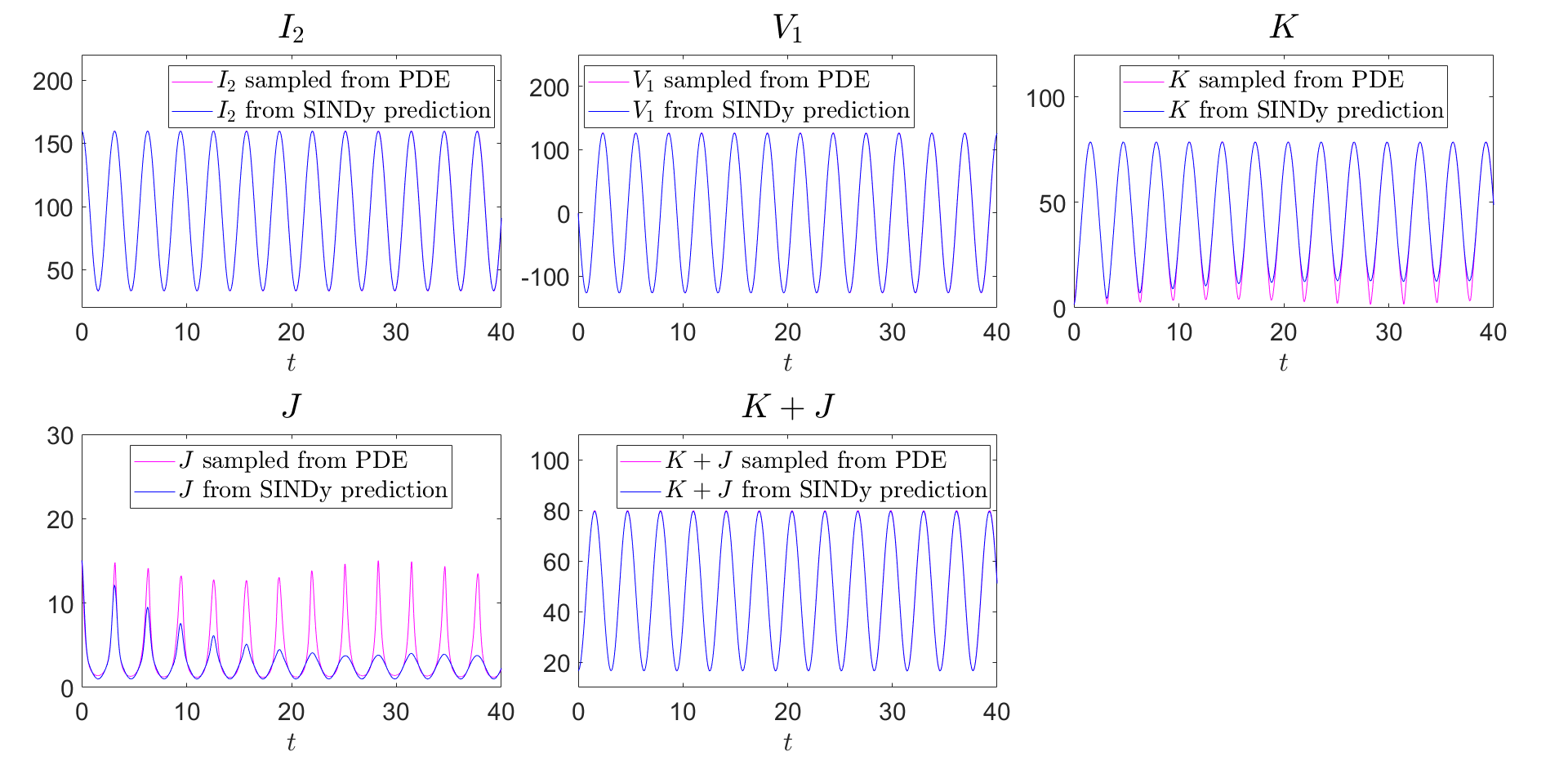}
\end{center}
\caption{Comparison of the ground-truth and SINDy-predicted time evolutions of \([I_2, V_1, K, J, K+J]\). SINDy is trained only on the selected moments \(\bfx = [I_2, V_1, K, J]\), where a closure does not exist, using a quadratic library \(\mathbf{\Theta}_{\deg\le 2}(\bfx)\). Similar to Figure~\ref{fig:ex_3_sindy_lib_1}, SINDy sacrifices accurate prediction for \(K\) and \(J\) \textit{individually} but indirectly captures the correct time evolution of \(E=K+J\).}\label{fig:ex_3_sindy_lib_2}
\end{figure}

Figure~\ref{fig:ex_3_sindy_lib_2} compares the ground-truth time evolutions of $[I_2, V_1, K, J, E = K+J]$ obtained from PDE integration with their SINDy-predicted dynamics after integrating  Eqs.~\eqref{eq:ex_3_sindy_raw_lib_2} and Eq.~\eqref{eq:ex_3_sindy_transformed_lib_2}. Similar to Figure~\ref{fig:ex_3_sindy_lib_1}, SINDy with the quadratic library $\mathbf{\Theta}_{\deg\le 2}(\bfx)$ applied to the unclosed moments $[I_2, V_1, K, J]$ sacrifices the exact recovery of the time evolution of $J$ and $K$, but indirectly captures the correct time evolution of $E=J+K$. However, unlike the linear library case in Section~\ref{sec:ex_3_sindy_lib_1}, SINDy with the quadratic library can yield the correct \textit{time evolution} of $E=K+J$ but \textit{fails} to uncover the correct \textit{governing equation} for $E$ [Eq.~\eqref{eq:ex_3_sindy_transformed_lib_2}] due to the expanded library size, leading to overfitting issues discussed above.

Here, we see a key complication within SINDy that has been also present in works such as~\citep{kutz2} and
especially~\citep{kutz1}, namely the methodology
is likely to result in ODE models that are
proximal to the theoretically expected ones,
but not identical to them. This, in turn,
may result in a nontrivial error outside
of the training set. Especially when the
library at hand is ``richer'' than the terms
expected to be present, unfortunately, it does
not generically seem that the reduced, theoretically
expected model will be discovered. Rather, our results
suggest that it is possible that
the additional ``wealth'' of the libraries used
can be leveraged to approximate the data via
different (nonlinear) dependent variable combinations.

As an even more problematic example, applying SINDy with a quadratic library to the moment time series generated from another IC \eqref{eq:ex_3_IC_4} results in a predicted ODE system that not only fails to match the ground truth, even after the theoretically suggested coordinate transformation $K+J$, but also causes the ODE system to blow up in finite time. For a detailed discussion, we refer the interested reader to the appendix.

\subsubsection{Stiefel optimization for discovering coordinate transformations}

Next, we test the methodology from Section~\ref{sec:stiefel_opt} to discover the coordinate transformation needed to close the moment system.

\noindent \textbf{Case 1: $\mu=0$.} We first consider the case where the hyperparamter $\mu$ in Eq.~\eqref{eq:optimization_inner_outer}is set to $\mu=0$, i.e., we do not promote sparsity in the matrix $\mathbf{\Xi}$. We use the linear library \(\mathbf{\Theta}(\bfx) = \mathbf{\Theta}_{\deg = 1}(\bfx)\), and set the maximum number of iterations to \(\texttt{maxIter} = 150\) in Algorithm~\ref{alg:inner_outer_optimization_new}. The algorithm produces the following outputs
\begin{align}
\label{eq:stifel_pred_sol}
    \mathbf{A}_{\text{out}}
    \approx
    \begin{bmatrix}
        -0.973 &  0.179 &   0.144 \\
        -0.169& -0.982&  0.083 \\
        -0.111& -0.040& -0.697\\
        -0.111 & -0.040 &-0.697
    \end{bmatrix}, \quad
    \mathbf{\Xi}_{\text{out}}
    \approx
    \begin{bmatrix}
        -0.024& -1.076&  0.005 \\
        1.016&  0.470& -0.540\\
        0.913&  5.778& -0.446
    \end{bmatrix}.
\end{align}
The coefficients are rounded to three decimal places. On the other hand, the ``ground-truth" coordinate transformation $\mathbf{A}_{\text{gt}}$ (after normalization to satisfy the Stiefel manifold constraint) and the corresponding dynamics $\mathbf{A}_{\text{gt}}$ according to Eq.~\eqref{eq:third_ME} are:
\begin{align}
\label{eq:stifel_gt_sol}
    \hspace{-2em}\mathbf{A}_{\text{gt}}
    =
    \begin{bmatrix}
        1 & 0 & 0\\
        0 & 1 & 0 \\
        0 & 0 & \frac{1}{\sqrt{2}}\\
        0 & 0 & \frac{1}{\sqrt{2}}
    \end{bmatrix}
    \approx
        \begin{bmatrix}
        1 & 0 & 0\\
        0 & 1 & 0 \\
        0 & 0 & 0.707\\
        0 & 0 & 0.707
    \end{bmatrix},~~
    \mathbf{\Xi}_{\text{gt}}
    =
    \begin{bmatrix}
        0 & -2 & 0\\
        1 & 0 & -\frac{1}{2\sqrt{2}}\\
        0 & 4\sqrt{2} & 0
    \end{bmatrix}
    \approx
        \begin{bmatrix}
        0 & -2 & 0\\
        1 & 0 & -0.354\\
        0 & 5.657 & 0
    \end{bmatrix}
\end{align}
At first glance, the predicted solution Eq.~\eqref{eq:stifel_pred_sol} appears different from the ground truth in Eq.~\eqref{eq:stifel_gt_sol}. However, after a further change of coordinates using
\begin{align}
    \mathbf{O} =
    \begin{bmatrix}
        -0.973  & -0.169 & -0.156\\
        0.179 & -0.982 & -0.0565\\
        0.144 & 0.0830 & -0.986
    \end{bmatrix}\in \text{O}(3),
\end{align}
we have
\begin{align}
    \mathbf{A}_{\text{out}}\mathbf{O}
    =
    \begin{bmatrix}
        1.000  & -0.000 & 0.000\\
        0.000 & 1.000 &  -0.000\\
        0.000 &  -0.000  &  0.707\\
       0.000 &   0.000  &  0.707\\
    \end{bmatrix}
    \approx \mathbf{A}_{\text{gt}}, ~~
    \mathbf{O}^\top \mathbf{\Xi}_{\text{out}} \mathbf{O}
    =
    \begin{bmatrix}
        0.000  & -2.000  &  0.000\\
        1.000  &  0.000  & -0.354\\
        0.000  &  5.657  &  0.000
    \end{bmatrix}
    \approx \mathbf{\Xi}_{\text{gt}}.
\end{align}
Hence, according to Remark~\ref{rmk:rmk_2}, $(\mathbf{A}_{\text{out}}, \mathbf{\Xi}_{\text{out}})$ and $(\mathbf{A}_{\text{gt}}, \mathbf{\Xi}_{\text{gt}})$ are equivalent solutions, and our method successfully predicts the correct coordinate transformation $\mathbf{A}_{\text{out}}$ to close the moment system. However, as expected, due to $\mu=0$, the linear combination matrix $\mathbf{\Xi}_{\text{out}}$ of the transformed dictionary is not sparse.

\noindent \textbf{Case 2: $\mu > 0$.} We next explore the case when $\mu \neq 0$ in Eq.~\eqref{eq:optimization_inner_outer}. We use the same linear library \(\mathbf{\Theta}(\bfx) = \mathbf{\Theta}_{\deg = 1}(\bfx)\), and $\mu$ is initially set to $\mu=1$. In Algorithm~\ref{alg:inner_outer_optimization_new}, we set 
\(\texttt{maxIter} = 1000\),  \(\texttt{Iter\_scheduled} = 400\), and \({\alpha} = 20\). The algorithm returns

\begin{align}
\label{eq:stifel_pred_sol_lasso}
    \widetilde{\mathbf{A}}_{\text{out}}
    \approx
    \begin{bmatrix}
        -1.000 &  0.000 &   0.030 \\
        0.000& -0.100&  0.002 \\
        -0.021& -0.001& -0.707\\
        -0.021 & -0.001 &-0.707
    \end{bmatrix}, \quad
    \widetilde{\mathbf{\Xi}}_{\text{out}}
    \approx
    \begin{bmatrix}
        0.000& -1.830&  0.003 \\
        1.000&  0.011& -0.383\\
        0.000&  5.714& -0.010
    \end{bmatrix}.
\end{align}
Similarly, after a further change of coordinates using:
\begin{align}
    \label{trans_matrix_lasso}
    \widetilde{\mathbf{O}} =
    \begin{bmatrix}
        -1.000  & 0.000 & -0.030\\
        0.000 & -1.000 & -0.002\\
        0.030 & 0.002 & -1
    \end{bmatrix}\in \text{O}(3),
\end{align}
we again have
\begin{align}
    \widetilde{\mathbf{A}}_{\text{out}}\widetilde{\mathbf{O}}
    =
    \begin{bmatrix}
        1.000  & -0.000 & 0.000\\
        0.000 & 1.000 &  -0.000\\
        0.000 &  -0.000  &  0.707\\
       0.000 &   0.000  &  0.707\\
    \end{bmatrix}
    \approx \mathbf{A}_{\text{gt}}, ~~
    \widetilde{\mathbf{O}}^\top \widetilde{\mathbf{\Xi}}_{\text{out}} \widetilde{\mathbf{O}}
    =
    \begin{bmatrix}
        0.000  & -2.000  &  0.000\\
        1.000  &  0.000  & -0.354\\
        0.000  &  5.657  &  0.000
    \end{bmatrix}
    \approx \mathbf{\Xi}_{\text{gt}}.
\end{align}

Note that the result $(\widetilde{\mathbf{A}}_{\text{out}}, \widetilde{\mathbf{\Xi}}_{\text{out}})$ in Eq.~\eqref{eq:stifel_pred_sol_lasso} for $\mu>0$ is much closer to the ground truth $(\mathbf{A}_{\text{gt}}, \mathbf{\Xi}_{\text{gt}})$ in Eq.~\eqref{eq:stifel_gt_sol}, compared to  $(\mathbf{A}_{\text{out}}, \mathbf{\Xi}_{\text{out}})$ in Eq.~\eqref{eq:stifel_pred_sol} with $\mu=0$.  This demonstrates that a positive $\mu$ with annealed optimization not only discovers the correct coordinate transformation to close the system but also achieves a sparser solution for $\mathbf{\Xi}$, reducing the number of terms on the right-hand side of the moment ODE system.

\subsection{An example of unclosed moment system}

Finally, we present a case without analytical closure. Our ``data-driven closure'' aims to provide an accurate, reduced-order description of the PDE by approximating the evolution of the moment systems. In principle,
this is the type of problem that we are aiming for,
namely the discovery of potential moment closures when
these may not be analytically available; the examples
presented previously are valuable benchmarks to 
raise the complications that may emerge when one
seeks to use this type of methodology in systems where
the answer may be unknown and what credibility one may
wish to assign to the obtained results.

Specifically, we consider an NLS equation~\eqref{eq:nls_harmonic} with a \textit{time-dependent} nonlinearity:
\begin{align}
\label{eq:unclosed_nonlinearity}
    g(\rho, t) = \left(\sin(t) + 2\right)\left|\rho\right|^{2}.
\end{align}
Notice that such time-dependent nonlinearities are well-known for some time
in atomic physics settings~\citep{donley,staliunas,FRM}
(and continue to yield novel insights to this day~\citep{autores})
and similar dynamical scenarios have
been considered in nonlinear optics~\citep{psaltis1,psalt2}.

Moment systems with such nonlinearity will not close to the best of our knowledge, so we aim to numerically approximate form of the dynamics of the moments \(\bfx = [I_2, V_1, E]\), where $E = K+J$. We consider the following three ICs:
\begin{align}
    u_1(x,0) &= 1.88\exp\left(-\frac{1}{2}\left(x-5\right)^{2}\right),
    \label{eq:unclosed_IC1}\\
    u_2(x,0) &= 1.88\left(\cos(2x) + \sin(2x)\right)\exp(-x^{2}),
    \label{eq:unclosed_IC2}\\
    u_3(x,0) &= \exp\left(-0.1x^{2}\right)\exp\left(0.1ix^{2}\right).
    \label{eq:unclosed_IC3}
\end{align}
Notably, the IC~\eqref{eq:unclosed_IC3}  includes a quadratic phase, motivated by the quadratic phase approximation (QPA) ansatz for NLS equations discussed by \citep{victor_theory}.
It also includes  regular, smooth localized initial
conditions, as well as one involving Fourier mode
oscillations, modulated by the Gaussian term.
We expect the SINDy-predicted dynamics to vary with the different ICs \eqref{eq:unclosed_IC1}, \eqref{eq:unclosed_IC2}, and \eqref{eq:unclosed_IC3},
hence the relevant choices.

Due to the periodic nature of the nonlinearity $g$ in Eq.~\eqref{eq:unclosed_nonlinearity}, we expect the moment system to be non-autonomous and exhibit an oscillatory pattern. To capture this, we introduce the following two ``auxiliary moments'':
\begin{align}
\left\{
\begin{aligned}
    H &= \sin(t) + 2,\\
    Q &= \cos(t) + 2.
\end{aligned}\right.
\end{align}
Naturally, one can observe that these are inspired
by the nature of $g(\rho,t)$. However, one can envision
the use of Fourier modes even if the mathematical 
model was not
known or if the data stemmed from experimental observations.

We then apply SINDy with a linear library $\mathbf{\Theta}_{\deg = 1}(\widetilde{\bfx})$ to the expanded moment systems $\widetilde{\bfx} = [\bfx, H, Q] = [I_2, V_1, E, H, Q]$.  The time series data of the moments are collected by integrating the PDE up to $T = 20$. The exact SINDy-predicted ODEs for all three ICs are presented in the appendix.

\begin{figure}[h!]
\begin{center}
\includegraphics[width=0.85\linewidth]{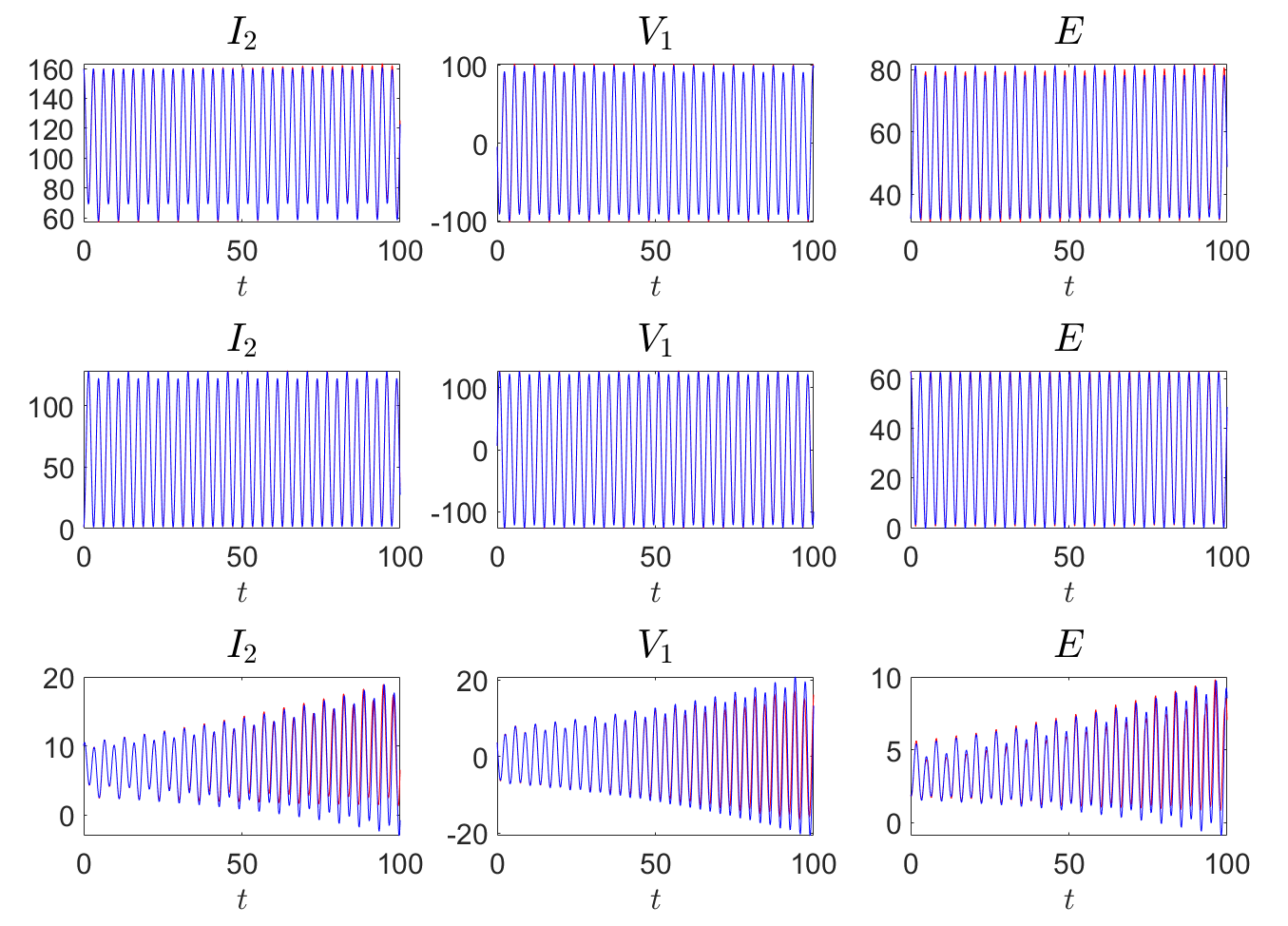}
\end{center}
\caption{Comparison between the ground-truth time evolution (in red) and the SINDy-predicted evolution (in blue). Training data were obtained by integrating the NLS equation up to \(T=20\) using the initial conditions \eqref{eq:unclosed_IC1}, \eqref{eq:unclosed_IC2}, and \eqref{eq:unclosed_IC3}. In all cases (displayed in three different rows), the SINDy-predicted moment dynamics closely match the ground truth, 
well past the time frame of the training data,
offering a satisfactory reduced-order description of the underlying PDE through the corresponding moment systems.}\label{fig:unclosed}
\end{figure}

To test the accuracy of the predicted moment systems, we integrate the SINDy-predicted ODEs into the future, up to $T=100$. Figure~\ref{fig:unclosed} compares the ground-truth and SINDy-predicted time evolutions for the three ICs \eqref{eq:unclosed_IC1}, \eqref{eq:unclosed_IC2}, and \eqref{eq:unclosed_IC3}. In all cases, the model closely matches the ground truth up to $T=100$. Indeed, this
is well past the training time of $T=20$ and thus
provides a satisfactory reduced-order description of the underlying PDE effective moment dynamics. Thus, despite
the potential shortcomings of the method which we 
tried to present in an unbiased fashion in 
case examples where the analytical theory helps assess
them, we still find it to be a worthwhile tool to consider.
I.e., from a data-driven perspective, it can be seen 
to potentially provide an
effective, low-dimensional dynamical representation
of the associated high-dimensional PDE dynamics.

\section{Conclusion and Future Directions}
This study explores a data-driven approach to identifying moment equations in nonlinear Schrödinger  models. 
This paves the way more generally towards the use
of similar methods in nonlinear PDEs which may feature
similar wave phenomena.
We applied the relevant sparse regression/optimization
methodology aiming to rediscover known analytical 
closures (progressively extending considerations to
more complex settings),
addressed overfitting by augmenting datasets with multiple initial conditions, and identified necessary coordinate transformations for systems requiring them so as to
bring forth the reduced or analytically tractable form
of the dynamics. Additionally, we demonstrated that our approach could provide a reduced-order description of systems without analytical closures by approximating the evolution of the moment systems. Our findings show that this data-driven method can capture complex dynamics in NLS models and offer insights for various physical applications, possibly well past the training time
used for the data-driven methods.

Future work will focus on extending our method to more complex PDEs and exploring its applicability to other types of nonlinear dynamical high-dimensional models,
such as, e.g., the ones we mentioned in the
context of~\citep{BELMONTEBEITIA20143267}. 
Another possible avenue is to, instead of recovering the moment systems through \textit{numerical differentiation} of the moment time series (as done in SINDy), leverage \textit{numerical integration} into future time
of a suitably augmented system. This approach, similar to Neural ODEs \citep{chen2018neural} and shooting methods, can help avoid producing predicted ODEs that blow up in finite time which SINDy may produce (see details in Section~\ref{sec:blowup} and the
Supplemental Material).
Additionally, we plan to develop techniques to identify \textit{nonlinear} coordinate transformations that can close the moment system, further enhancing the applicability of our method. Lastly, one can envision
such classes of techniques for obtaining additional
reduced features of solitary waves, such as data-driven
variants of the variational approximation~\citep{malomed2002},
or data-driven models of soliton interaction
dynamics~\citep{MANTON1979397,Manton2,ricardo}.

\section*{Acknowledgement}
This material is based upon work supported by the U.S. National Science Foundation under the awards PHY-2110030 and DMS-2204702 (PGK), as well as DMS-2052525, DMS-2140982, and DMS2244976 (WZ).

\bibliographystyle{abbrvnat} 
\bibliography{main.bib}

\newpage
\appendix
\onecolumn

In this appendix, we provide additional experimental results that complement the ones presented in the main text.

\appendix

\section{Examples with analytical moment closure}

\subsection{Example 1}

We present here the negative result mentioned in the main text. Recall that the chosen moments are \(\bfx = [I_1, V_0]\), and we consider the following two initial conditions (ICs):
\begin{align}
    u_1(x,0) &= \pi^{-1/4}\exp\left(-\frac{1}{2}\left(x-5\right)^{2}\right),
    \label{first IC for Harmonic O}\\
    u_2(x,0) &= \frac{1}{2}\text{sech}^{2}\left(x-5\right),
    \label{second IC for Harmonic O}
\end{align}
When we apply SINDy with a cubic library \(\mathbf{\Theta}_{\deg \le 3}(\bfx)\) to the moment time series data generated \textit{only} from IC~\eqref{first IC for Harmonic O}, SINDy erroneously predicts the wrong dynamics:
\begin{equation}\label{eq: Negative example}
\left\{
    \begin{aligned}
        \frac{dI_1}{dt} &= 0.040I_1^{2}V_0 + 0.040V_0^{3},\\
        \frac{dV_0}{dt} &= -0.040I_1^{3} - 0.040I_1V_0^{2}.
    \end{aligned}\right.
\end{equation}
Note that this issue can be resolved if the data from both ICs are combined before applying SINDy, resulting in the correct ODE system (harmonic oscillator) presented in the main text.

\subsection{Example 2}

Let the selected moments be $\bfx=[I_2, V_1, K]$, and we consider the same ICs \eqref{first IC for Harmonic O} and \eqref{second IC for Harmonic O} as before.

\noindent \textbf{Negative result when a \textit{single} IC is used.} If we apply SINDy with a quadratic library $\mathbf{\Theta}_{\deg \le 2}(\bfx)$ to the moment data from a single IC~\eqref{first IC for Harmonic O}, we have
\begin{align}
    \left\{
    \begin{aligned}
        &\frac{dI_2}{dt} = 0.038I_2V_1 + 0.077V_1K\\
        &\frac{dV_1}{dt} = -0.077 I_2^2 + 0.308 K^2\\
        &\frac{dK}{dt} = -0.062V_1K
    \end{aligned}\right.
\end{align}

\noindent \textbf{Positive result when \textit{both} ICs are combined.} On the other hand, if we apply SINDy with a quadratic library $\mathbf{\Theta}_{\deg \le 2}(\bfx)$ to the moment data from two ICs \eqref{first IC for Harmonic O} and \eqref{second IC for Harmonic O} combined, we have
\begin{align}
    \left\{
    \begin{aligned}
        &\frac{dI_2}{dt} = 1.000 V_1\\
        &\frac{dV_1}{dt} = -2.000I_2+4.000K\\
        &\frac{dK}{dt} = -0.500V_1
    \end{aligned}\right.
\end{align}

\section{Examples where closure exists after coordinate transformations}

Recall that in this example, the selected moments are $\bfx = [I_2, V_1, K, J]$, and the moment system only closes after a coordinate transformation, such as $E = K+J$. We consider the following four distinct ICs, 
\begin{align}
    u_1(x,0) &= \pi^{-1/4}\exp\left(-\frac{1}{2}\left(x-5\right)^{2}\right),
    \label{IC1 for dynamics 3}\\
    u_2(x,0) &= 1.88\exp\left(-\frac{1}{2}\left(x-5\right)^{2}\right),
    \label{IC2 for dynamics 3}\\
    u_3(x,0) &= 1.88\left(\exp\left(-\frac{1}{2}\left(x-5\right)^{2}\right) + \exp\left(-\left(x-2\right)^{2}\right)\right),
    \label{IC3 for dynamics 3}\\
    u_4(x,0) &=  1.88\left(\cos(2x) + \sin(2x)\right)\exp(-x^{2}).
    \label{IC4 for dynamics 3}
\end{align}
We use SINDy with different library sizes directly on the selected moments \(\bfx = [I_2, V_1, K, J]\), where a closure does not exist.

\subsection{SINDy with linear library $\mathbf{\Theta}_{\deg=1}(\bfx)$}

\begin{figure}[h!]
\begin{center}
\includegraphics[width=0.85\linewidth]{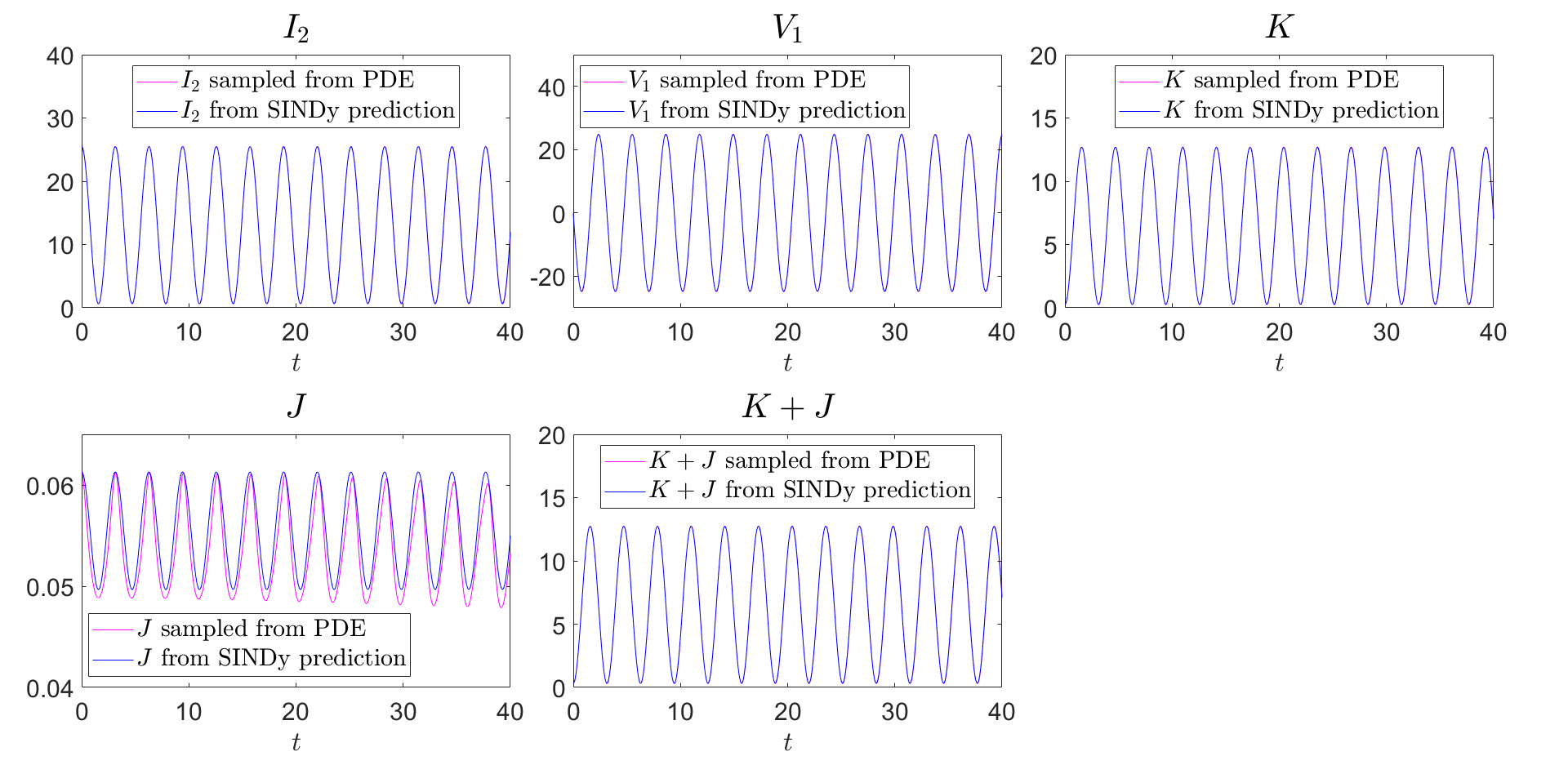}
\end{center}
\caption{Comparison of the ground-truth and SINDy-predicted time evolutions of \([I_2, V_1, K, J, K+J]\), trained with the IC \eqref{IC1 for dynamics 3}.}
\label{fig:ex_3_IC_1}
\end{figure}

\begin{figure}[h!]
\begin{center}
\includegraphics[width=0.85\linewidth]{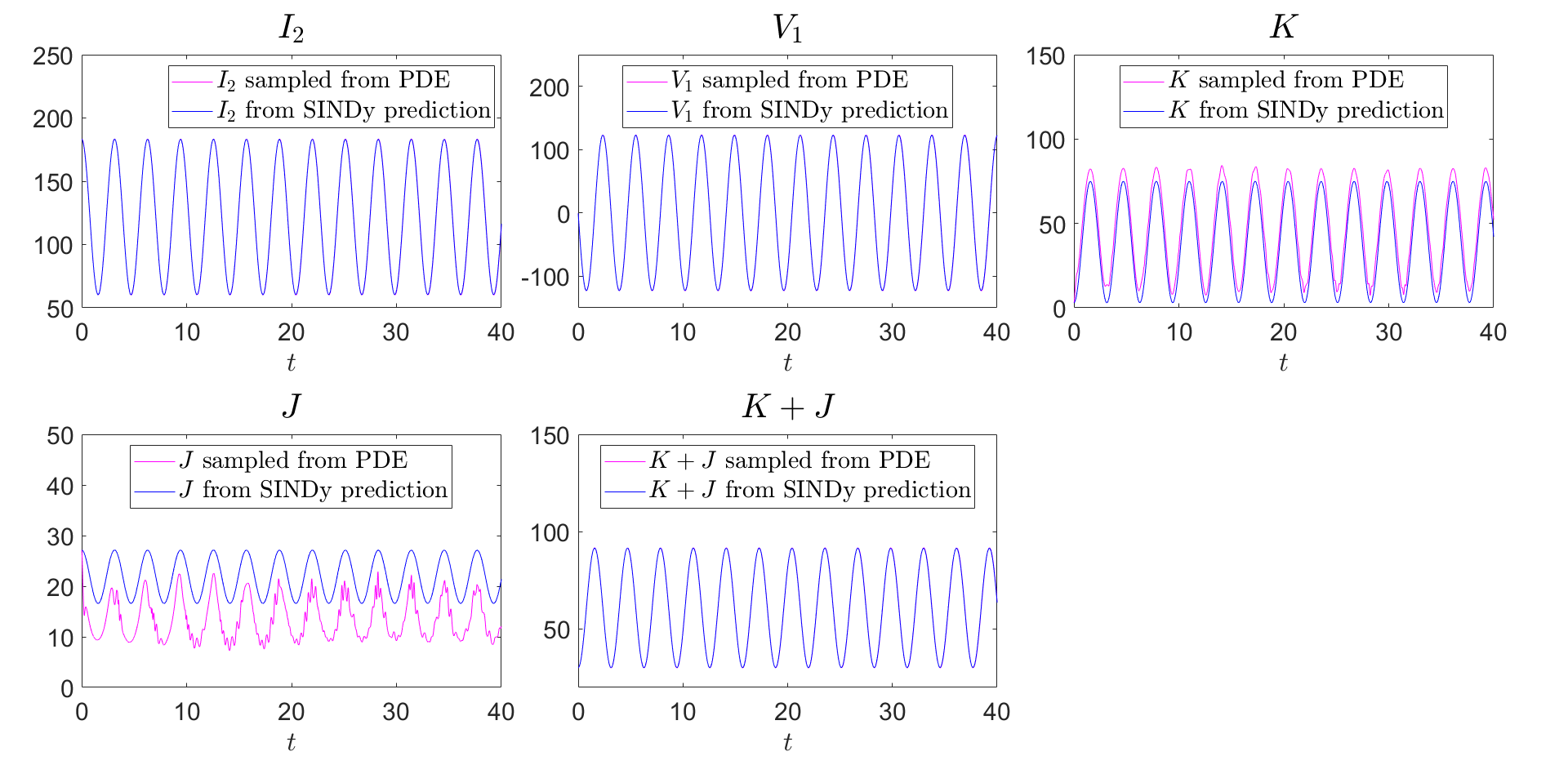}
\end{center}
\caption{Comparison of the ground-truth and SINDy-predicted time evolutions of \([I_2, V_1, K, J, K+J]\), trained with the IC \eqref{IC3 for dynamics 3}.}
\label{fig:ex_3_IC_3}
\end{figure}

\begin{figure}[h!]
\begin{center}
\includegraphics[width=0.85\linewidth]{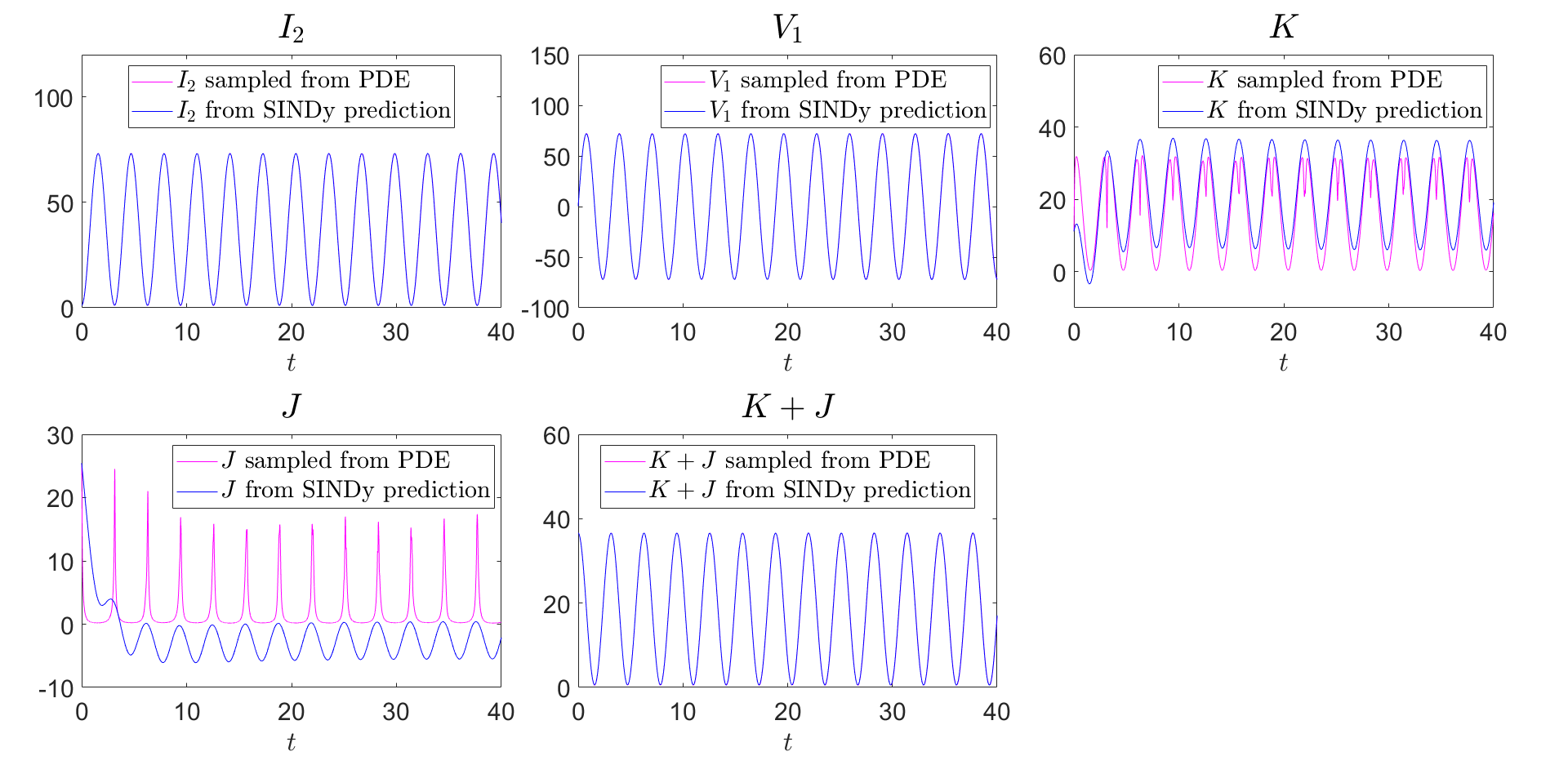}
\end{center}
\caption{Comparison of the ground-truth and SINDy-predicted time evolutions of \([I_2, V_1, K, J, K+J]\), trained with the IC \eqref{IC4 for dynamics 3}.}
\label{fig:ex_3_IC_4}
\end{figure}

The result of applying SINDy with a lineary library, $\mathbf{\Theta}_{\deg=1}(\bfx)$, to moment time series data generated from IC~\eqref{IC2 for dynamics 3} is already shown in the main text. Here, we present the results for the other three ICs.
\begin{itemize}
    \item For the IC~\eqref{IC1 for dynamics 3}, the SINDy-predicted ODEs are:
    \begin{align}
    \left\{
    \begin{aligned}
        \frac{dI_2}{dt} &= 1.000V_1, \\
        \frac{dV_1}{dt} &= -1.991I_2 + 4.015K, \\
        \frac{dK}{dt} &= -0.500V_1,\\
        \frac{dJ}{dt} &= 0.000V_1,       
    \end{aligned}\right.
    \end{align}
    \item For the IC~\eqref{IC3 for dynamics 3}, the SINDy-predicted ODEs are:
    \begin{align}
    \left\{
    \begin{aligned}
        \frac{dI_2}{dt} &= 1.000V_1,\\
        \frac{dV_1}{dt} &= -2.000I_2 + 4.000K + 3.998J,\\
        \frac{dK}{dt} &= -0.586V_1,\\
        \frac{dJ}{dt} &= 0.086V_1.
    \end{aligned}\right.
    \end{align}
    \item For the IC~\eqref{IC4 for dynamics 3}, the SINDy-predicted ODEs are:
    \begin{align}
    \left\{
    \begin{aligned}
        \frac{dI_2}{dt} &= 1.000V_1,\\
        \frac{dV_1}{dt} &= -2.000I_2 + 4.000K + 3.999J,\\
        \frac{dK}{dt} &= -2601 + 35.038I_2 - 0.419V_1 + 70.029K + 70.707J,\\
        \frac{dJ}{dt} &= 2601 - 35.037I_2 - 0.081V_1 - 70.028K - 70.705J.
    \end{aligned}\right.
\end{align}
\end{itemize}
Figures~\ref{fig:ex_3_IC_1}, \ref{fig:ex_3_IC_3}, and \ref{fig:ex_3_IC_4} respectively compare the ground-truth time evolutions of \([I_2, V_1, K, J, E=K+J]\) obtained from PDE integration with their SINDy-predicted dynamics under ICs \eqref{IC1 for dynamics 3}, \eqref{IC3 for dynamics 3}, and \eqref{IC4 for dynamics 3}.

\subsection{SINDy with quadratic library $\mathbf{\Theta}_{\deg \le 2}(\bfx)$}

When SINDy with a quadratic library is applied to moments $[I_2, V_1, K, J]$ generated from the IC~\eqref{IC4 for dynamics 3}, we obtain
\begin{align}
    \left\{
    \begin{aligned}
        \frac{dI_2}{dt} =& 1.000V_1,\\
        \frac{dV_1}{dt} =& -1.983I_2 + 3.967K + 3.971J - 0.000I_2^{2} + 0.001K^{2} + 0.002KJ + 0.001J^{2},\\
        \frac{dK}{dt} = & 7129180.673 -191977.284I_2 -42.858V_1 - 384146.774K -385838.007J \\
        &+ 1292.761I_2^{2} + 0.571I_2V_1 + 5125.234I_2K + 5149.187I_2J + 6.050V_1^{2} \\
        &+ 1.124V_1K + 1.675V_1J + 5176.223K^{2} + 10398.013KJ + 5221.762J^{2},\\
        \frac{dJ}{dt} =& -7129455.326 + 191984.680I_2 + 42.359V_1 + 384161.569K + 385852.833J\\ &-1292.811I_2^{2} -0.571I_2V_1 -5125.433I_2K -5149.386I_2J-6.050V_1^{2}\\
        &-1.124V_1K -1.676V_1J-5176.422K^{2} -10398.412KJ -5221.962J^{2}.
    \end{aligned}\right.
\end{align}

This predicted ODE system poses two major concerns. First, adding the equations of \(K\) and \(J\) no longer recovers the ground-truth equation for $E$, as, e.g., the constant terms do not cancel each other out. Second, the solution blows up in finite time (e.g., \(t = 5\)), unlike the ground-truth dynamics, which are defined for all $t>0$. In summary, this indicates that SINDy with a quadratic library fails to discover the underlying closed moment system.

\section{An example of unclosed moment system}

Recall that for the unclosed moment system associated with an NLS equation with a time-dependent nonlinearity,
\begin{align}
    g(\rho, t)= \left(\sin(t)+2\right)|\rho|^2,
\end{align}
we considered the following three ICs:
\begin{align}
    u_1(x,0) &= 1.88\exp\left(-\frac{1}{2}\left(x-5\right)^{2}\right),
    \label{Gaussian IC again for time-dependent NLS}\\
    u_2(x,0) &= 1.88\left(\cos(2x) + \sin(2x)\right)\exp(-x^{2}),
    \label{Fourier-Mode IC again for the time-dependent NLS}\\
    u_3(x,0) &= \exp\left(-0.1x^{2}\right)\exp\left(0.1ix^{2}\right).
    \label{QPA type IC}
\end{align}
The SINDy-predicted ODEs are, respectively,
\begin{itemize}
    \item For IC \eqref{Gaussian IC again for time-dependent NLS},
    \begin{align}
        \left\{
        \begin{aligned}
            \frac{dI_2}{dt} &= 1.000V_1,\\
            \frac{dV_1}{dt} &= 22.162 -2.098I_2 + 3.799 E,\\
            \frac{dE}{dt} &= -8.687 -0.497V_1 + 4.343Q,\\
            \frac{dH}{dt} &= -2.001 + 1.000Q,\\
            \frac{dQ}{dt} &= 2.001 - 1.000H.
        \end{aligned}\right.
    \end{align}
    \item For IC \eqref{Fourier-Mode IC again for the time-dependent NLS},
    \begin{align}
        \left\{
        \begin{aligned}
            \frac{dI_2}{dt} &= 1.000V_1,\\
            \frac{dV_1}{dt} &= -1.999I_2 + 3.998E,\\
            \frac{dE}{dt} &= -4.405 -0.501V_1 + 2.202Q,\\
            \frac{dH}{dt} &= -2.001 + 1.000Q,\\
            \frac{dQ}{dt} &= 2.001 - 1.000H.   
        \end{aligned}\right.
    \end{align}
    \item For IC \eqref{QPA type IC},
    \begin{align}
        \left\{
        \begin{aligned}
            \frac{dI_2}{dt} &= 0.999V_1,\\
            \frac{dV_1}{dt} &= 6.979 -2.554I_2 + 2.927E + 0.030H + 0.284Q,\\
            \frac{dE}{dt} &= -1.659 -0.021I_2 -0.480V_1 + 0.023E -0.107H + 0.979Q,\\
            \frac{dH}{dt} &= -2.001 + 1.000Q,\\
            \frac{dQ}{dt} &= 2.001 - 1.000H. 
        \end{aligned}\right.
    \end{align}
\end{itemize}

\end{document}